\documentclass[twocolumn,showpacs,floats,floatfix,superscriptaddress,aps,pra]{revtex4-1}
\usepackage{amsfonts}
\usepackage{amssymb}
\usepackage{amsmath}
\usepackage{calc,graphicx,color,bm}
\usepackage{subfigure}
\usepackage{graphicx}
\usepackage{dcolumn}
\usepackage{bm}
\usepackage{color} 
\usepackage{CJK}

\newcommand{\beq}{\begin{equation}}
\newcommand{\eeq}{\end{equation}}
\newcommand{\beqa}{\begin{eqnarray}}
\newcommand{\eeqa}{\end{eqnarray}}

\usepackage{color}

\begin{document}

\title{Reverse engineering protocols for controlling spin dynamics}
\date{\today}

\author{Qi Zhang}
\affiliation{Department of Physics, Shanghai University, 200444
Shanghai, People's Republic of China}
\affiliation{Universit\'{e} de Toulouse, UPS, Laboratoire Collisions Agr\'{e}gats R\'{e}activit\'{e}, IRSAMC, F-31062 Toulouse, France}
\affiliation{CNRS, UMR 5589, F-31062 Toulouse, France}

\author{Xi Chen}
\email{xchen@shu.edu.cn}
\affiliation{Department of Physics, Shanghai University, 200444 Shanghai, People's Republic of China}

\author{D. Gu\'ery-Odelin}
\email{dgo@irsamc.ups-tlse.fr}
\affiliation{Universit\'{e} de Toulouse, UPS, Laboratoire Collisions Agr\'{e}gats R\'{e}activit\'{e}, IRSAMC, F-31062 Toulouse, France}
\affiliation{CNRS, UMR 5589, F-31062 Toulouse, France}

\begin{abstract}
We put forward reverse engineering protocols to shape in time the components of the magnetic field to manipulate a single spin, two independent spins with different gyromagnetic factors, and two interacting spins in short amount of times.
We also use these techniques to setup protocols robust against the exact knowledge of the gyromagnetic factors for the one spin problem, or to generate entangled states for two or more spins coupled by dipole-dipole interactions.
\end{abstract}
\maketitle

\section{Introduction}
The implementation of quantum computing and quantum information processing require a careful preparation of the initial quantum state and accurate control of its further evolution in time \cite{BookLoss,BookShore}.
There is a large body of literature dealing with coherent control in quantum structures based on the precise tailoring of adiabatic pulses \cite{Bergmann-Rev1,Shapiro-Rev,Bergmann-Rev2} and pulse sequences \cite{nmr,Torosov-PRL}, or, alternatively, on the application of optimal control theory \cite{Tannor,Boscain,Albertini,Gerhard} to such problems. Some of these techniques generate solutions with sharp variations of the parameters, which may therefore pose a problem of practical implementation because of the finite time variation of parameters which one can afford on an experiment.

We propose here another strategy inspired by the reverse engineering protocols applied recently to the fast transport or manipulation of wave functions \cite{PRA2011,PRA2014,PRA2016NL,PRA2016W} or in tailored transformations in statistical physics \cite{Boltzmann,NaturePhys,APL}. Such techniques have emerged in the broader context of ``Shortcuts to Adiabaticity" (STA) \cite{RevueSTA,PRX}. The shortcut protocol consists in imposing the desired evolution of the dynamical quantity of interest and inferring from it the time evolution of the parameters. This provides an efficient way to reverse engineer an analytically solvable Schr\"{o}dinger equation for a
driven spin-1/2 system \cite{PRLAo,YuePRL,XiarXiv}, or, equivalently, a generic two-level system with cold atoms \cite{2012njp,Stephane,PRALu,SarmaPRL,Shore}. In contrast with the methods mentioned above, it can be stated so to generate smooth variations of the parameters in time (see \cite{Sun}).

The shaping of the three components of the time-dependent magnetic field that one shall apply to induce an arbitrary trajectory on the Bloch sphere of a single spin 1/2 has been explicitly worked out in Ref.~\cite{Berry09}. Interestingly, the equation of motion of the mean value of the spin cannot be reversed in an unique manner. It means in practice that there is a lot of freedom to reach a given target state and to fulfill also some extra requirements. We will take advantage of this feature in the following. In this article, we setup (i) a few general procedures for reverse engineering, (ii) an algorithm to build up the smooth variations in time of the magnetic field components that one should apply to spin flip a spin 1/2 (or connect two points on the Bloch sphere) in an arbitrary short amount of time, (iii) expand the parameter space of those solutions to fulfill extra requirements such as the robustness of the operation or the application of the transformation to two spins with different coupling strength to the magnetic field. We then discuss how those protocols shall be modified to take into account interactions between spins, and generate in an optimal amount of time entangled states of two or more spins.

\section{The reverse engineering protocols for a single spin}

In the reverse protocol, the magnetic field components as a function of time are deduced from the equations of evolution of the spin that is imposed according to the desired boundary values.
In practice, it can be useful to have different formulations of the same problem since the inversion \cite{PRAChen} or the generalization to higher dimension can be easier for one of them. Those ideas have a wide range of possible applications in various systems, including spin system \cite{Takahashi,Suter,Tokatly}, Bose-Einstein condensates (BECs) \cite{Oliver,Wu} and other many-body systems \cite{YAChen,campo1,campo2}. 

Hereafter, we propose to work out such an inversion with three different formulations of a spin 1/2 in a time-dependent magnetic field: (i) the direct reversing of the time evolution operator, (ii) the inversion of Modeling representation formulation of the problem and (iii) the inversion of the precession equations. This is a non exhaustive list. For instance, another common method for reverse engineering relies on dynamical invariant \cite{2012njp}, and will be used in Sec.~\ref{anis}.

\subsection{Inverting the time evolution operator}
\label{secexact}

We consider a spin 1/2 in an initial state $|\psi(0)\rangle$. Its time evolution is encapsulated in the evolution operator $U(t)$: $|\psi(t)\rangle=U(t)|\psi(0)\rangle$.
The most general form for $U$ is a 2$\times$2 complex matrix whose coefficients are partially related to ensure its unitary property. Denoting $u_{ij}=\rho_{ij} e^{i\varphi_{ij}}$, the coefficients of $U$ shall fulfill the following relations $\varphi_{12}=\varphi_{21}\equiv \varphi$, $\rho_{12}=\rho_{21}\equiv \rho$, $\rho_{11}=\rho_{22}\equiv r$ and $\varphi_{11}+\varphi_{22}=2 \varphi+\pi$. The most general form of  $U$ therefore reads
\begin{equation}
U(t)=\left(
\begin{array}{ll}
re^{i\varphi_{11}} & \sqrt{1-r^2}e^{i\varphi} \\
\sqrt{1-r^2}e^{i\varphi}  & re^{i(2\varphi+\pi-\varphi_{11})}
\end{array}
\right),
\end{equation}
where the variables $r$, $\varphi_{11}$ and $\varphi$ are time-dependent.
The reverse engineering protocol consists in shaping in time the three variables $r$, $\varphi_{11}$ and $\varphi$ to
ensure the transformation $|\psi(0)\rangle \longrightarrow |\psi(t_f)\rangle = |\psi_{\rm target}\rangle$.
The Schr\"{o}dinger equation $i\hbar \partial_t | \psi(t)\rangle = H | \psi(t)\rangle$ implies that $H=i \hbar \dot U U^\dagger$.
The expansion of $H$ on the Pauli matrices $\sigma_i$ ($i=x,y,z$) gives the time-dependent magnetic field components that should be implemented in order to
follow the desired trajectory: $H = -\gamma_1 {\bf s}_1\cdot {\bf B} (t)=-\gamma_1 \hbar {\boldsymbol \sigma}\cdot {\bf B}(t)/2$ with
\begin{eqnarray}
B_x (t)& = &   - \frac{2}{\gamma_1}  \frac{  r(1-r^2)\dot \Phi\cos\Phi+\dot r \sin \Phi}{\sqrt{1-r^2}}, \nonumber \\
B_y (t)& = &  - \frac{2}{\gamma_1} \frac{  r(1-r^2)\dot \Phi\sin\Phi-\dot r \cos \Phi}{\sqrt{1-r^2}}, \nonumber \\
B_z (t)& = &  - \frac{2}{\gamma_1} \dot \Phi r^2, \nonumber
\end{eqnarray}
where $\gamma_1$ is the gyromagnetic factor  and $\Phi=\varphi-\varphi_{11}$ the relative phase. We conclude that only two parameters are relevant in this case $r(t)$ and $\Phi(t)$.
This form generally contains the particular results deduced from other methods such as the tracking transitionless algorithm and the dynamical invariant approach \cite{RevueSTA}, see also Refs. \cite{SarmaPRL,Shore}.

As a simple example, let's work out the simplest form of the magnetic field components to ensure the spin flip of the spin. The wave function reads $|\psi(t) \rangle = r(t)e^{i\varphi_{11}(t)}|+ \rangle + \sqrt{1-r^2(t)} e^{i\varphi(t)}| -\rangle$, where $| \pm \rangle$ are the eigenstates of $\sigma_z$ with eigenvalues $\pm 1$, corresponding to the spin up and down. To ensure the spin flip from $| + \rangle$ to $| - \rangle$ in an amount of time $t_f$, we need to fulfill the following boundary conditions, $r(0)=1$ and $r(t_f)=0$. To avoid the divergence of denominators, we choose $\Phi=0$ and $r(t)=\cos(\theta(t))$ with $\theta(0)=0$ and $\theta(t_f)=\pi/2$. We find $B_x=0$ and $B_z=0$, and a solution that contains the famous $\pi$-pulse solution with a constant magnetic field $B^0_y=\pi/(\gamma_1 t_f)$. The general method that we have worked out enables one to have any type of final state, including superposition of states.

\subsection{Inverting the Madelung representation formulation}

Another strategy consists in using an exact semiclassical approach based on the phase-modulus equations, commonly referred to as the Modeling representation.
To derive the corresponding set of coupled equations we start by introducing the general form $| \psi(t)\rangle=a(t) |+\rangle +b(t) |-\rangle$ into the Schr\"odinger equation in the presence of a time-dependent magnetic field:
\begin{eqnarray}
i \hbar \dot a=-\frac{\gamma_1}{2} \hbar \left( a B_z+b B_x-ib B_y \right),\label{e1}\\
i \hbar \dot b=-\frac{\gamma_1}{2} \hbar \left(a  B_x-b B_z+ia B_y \right).\label{e2}
\end{eqnarray}
Let's rewrite the coefficients $a$ and $b$ in modulus-phase representation $a(t)=\sqrt{n_a(t)} e^{i \varphi_a(t)}$, $b(t)=\sqrt{n_a(t)} e^{i \varphi_b(t)}$. The two previous complex equations can be recast as a Hamiltonian problem for the conjugate variables $(n_a,\varphi_a)$ and $(n_b,\varphi_b)$: $\dot n_i=-\partial_{\varphi_i} H$, $\dot \varphi_i=\partial_{n_i} H$ with ($i=a, b$).
It is convenient to introduce the relative variables $\Delta_n(t)=n_a(t)-n_b(t)$ and $\theta(t)=\varphi_a(t)-\varphi_b(t)$.
The expression of the Hamiltonian now reads
\beq
H= \gamma_1 B_z \Delta_n/2+ \sqrt{1-\Delta_n^2} (B_x\cos \theta-B_y\sin \theta),
\eeq
and the dynamics is given by the two scalar equations
\begin{eqnarray}
\dot \Delta_n &=& \gamma_1 \sqrt{1-\Delta_n^2} (B_x\sin\theta+B_y\cos\theta),\\
\dot\theta &=& \gamma_1 \left[B_z- \frac{\Delta_n}{\sqrt{1-\Delta_n^2}} (B_x\cos\theta-B_y\sin\theta) \right].~~~
\end{eqnarray}
From the evolution of $\Delta_n$ and $\theta$, we can infer the components of the magnetic field. For instance,
\begin{eqnarray}
B_x(t) &=& \frac{\dot \Delta_n}{\gamma_1 \sqrt{1-\Delta_n^2}\sin\theta}, \nonumber \\
B_y(t) &=& 0, \nonumber \\
B_z(t) &=& \frac{1}{\gamma_1}\left(\dot \theta+\frac{\dot \Delta_n\Delta_n}{1-\Delta_n^2}\frac{\cos \theta}{\sin\theta}\right). \nonumber
\end{eqnarray}

\subsection{Inverting the precessions equations}

Alternatively, we can work out the equations of motion for the mean value of the spin
\begin{equation}
\label{eqs1}
\frac{{\rm d} \langle {\bf s}_1 \rangle }{{\rm d}t}   =   \frac{1}{i\hbar}\langle [{\bf s}_1,H] \rangle = \gamma_1  \langle {\bf s}_1 \rangle \times {\bf B}(t).
\end{equation}
In the following, we note  ${\bf S}_1=2 \langle {\bf s}_1 \rangle/\hbar$ and use the spherical coordinates to describe the motion of the spin on the Bloch sphere: ${\bf S}_1(\sin\theta(t)\cos\varphi(t),\sin\theta(t)\sin\varphi(t),\cos\theta(t))$. To work out our reverse protocol, we calculate  the left-hand side of precession equations (\ref{eqs1})
\begin{eqnarray}
 \dot {S}_{1x} &=&   \dot \theta \cos \theta  \cos \varphi- \dot \varphi \sin \theta \sin \varphi, \label{2a}   \\
 \dot {S}_{1y} & =&  \dot \theta \cos \theta \sin \varphi +\dot \varphi \sin \theta \cos \varphi, \label{2b}   \\
 \dot {S}_{1z} & =&   - \dot \theta \sin \theta. \label{2c}
\end{eqnarray}
Combining Eqs.~(\ref{2a})-(\ref{2c}), we get
\begin{eqnarray}
\dot \theta &=&   \gamma_1 \left(B_{y} \cos \varphi - B_{x}  \sin \varphi \right),  \label{22a}  \\
\dot \varphi   &= &  \gamma_1 [ B_{z}-\cot \theta (B_{x}  \cos \varphi + B_{y} \sin \varphi)],
\label{22b}
\end{eqnarray}
from which we infer the expression of the transverse magnetic field components. With this set of equations we already obtain  a class of solution by setting $B_x=B_z=0$ and $\varphi=0$, we find $\dot \theta=\gamma_1 B_y$. The reverse engineering protocol consists here in choosing for $\theta(t)$ a function that obeys the boundary conditions $\theta(0)=0$ and $\theta(t_f)=\pi$, and to infer from it the expression for $B_y(t)$. We can readily recover here also the $\pi$-pulse solution.

It is interesting to let the possibility to shape any curve on the Bloch sphere \cite{Berry09}. For this purpose, we need non trivial dependence of both $\theta(t)$ and $\varphi(t)$. However, as suggested by Eqs.~(\ref{22a}) and (\ref{22b}), we can engineer only transverse magnetic field components and impose the variation of the longitudinal magnetic field component. This choice amounts to using explicitly the non uniqueness of the solution. The solution is then quite simple, we set the evolution of $\theta(t)$, $\varphi(t)$ and $B_{z}(t)$ according to our boundary conditions. We have to be careful since we need to avoid divergences. This means that we have to take care of the terms having a $\tan \theta$. This latter terms diverge for $\theta=\pi/2$, at time $t=t^*$ for which $\theta(t^*)=\pi/2$. To compensate for this divergence, we have to cancel also $\dot \varphi (t^*)=0$ and $B_{1z}(t^*)=0$. A way out for the last term consists in choosing $B_{1z}(t)=B_0\cos(\theta(t))$. The set of equations (\ref{22a}) and (\ref{22b}) then reads
\begin{eqnarray}
\label{b1xn}
B_{x} &=& -\frac{\dot \theta}{\gamma_1}\sin \varphi - \frac{\dot \varphi}{\gamma_1}\tan \theta \cos \varphi +B_0\sin \theta \cos \varphi,  \\
\label{b1yn}
B_{y} &=& \frac{\dot \theta}{\gamma_1}\cos \varphi - \frac{\dot \varphi}{\gamma_1}\tan \theta \sin \varphi +B_0\sin \theta \sin \varphi.
\end{eqnarray}

\begin{figure}[h!]
\begin{center}
\scalebox{0.7}[0.7]{\includegraphics{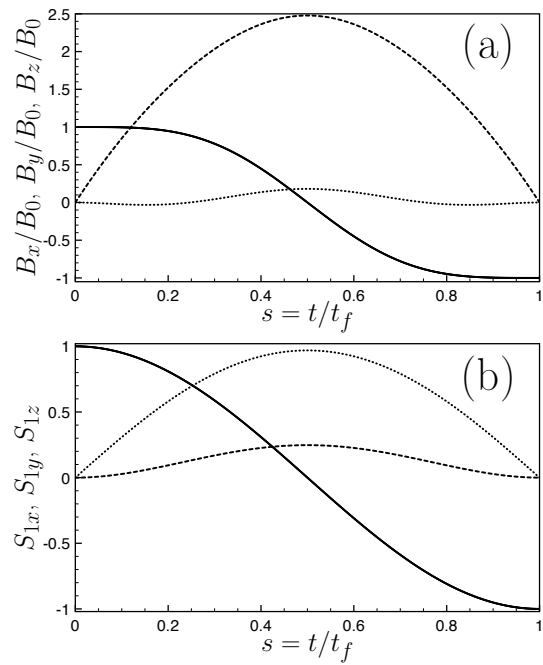}}
\caption{(a) Evolution of the magnetic field components $B_z/B_0$ (solid line), $B_y/B_0$ (dashed line) and $B_x/B_0$ (dotted line) as a function of time.
The shaping of the magnetic field components is obtained self-consistently from a reverse engineering protocol in which we impose the variations of the spin components according to the target state (spin flip in this case) and the time duration of the transformation. (b) Spin components $S_{1z}$ (solid line), $S_{1y}$ (dashed line) and $S_{1x}$ (dotted line) as a function of time. Parameters: $\tilde \gamma_1 (\equiv \gamma_1 t_f)=2$, $B_0=1$.}
\label{fig1}
\end{center}
\end{figure}

Consider the following example, we want to spin flip the spin from $|+\rangle$ to $|-\rangle$ in an amount of time $t=t_f$.
For convenience, we use in the following the dimensionless time $s=t/t_f$.
We use the boundary conditions $\theta(0)=0$ and $\theta(1)=\pi$. The simplest polynomial interpolation between those two boundary conditions is $\theta(s)=\pi s$. In this case, $s^*=1/2$. The boundary conditions for $\varphi$ are therefore $\varphi(0)=0$, $\varphi(1)=0$ and $\dot \varphi(1/2)=0$. We choose here a polynomial ansatz  $\varphi(s)=s-s^2$ to fulfill those conditions.
Equations (\ref{b1xn}) and (\ref{b1yn}) take then the simple form
\begin{eqnarray}
 B_{x} &=& -\frac{\pi }{\gamma_1 t_f}\sin (s-s^2) - \frac{1-2s}{\gamma_1 t_f}\tan (\pi s) \cos (s-s^2) \nonumber \\
 &+&B_0\sin(\pi s) \cos (s-s^2),~~~~~   \label{b1x} \\
 B_{y} &=& \frac{\pi  }{\gamma_1 t_f}\cos (s-s^2) - \frac{1-2s}{\gamma_1 t_f}\tan (\pi s) \sin (s-s^2) \nonumber \\
  &+&B_0\sin (\pi s) \sin (s-s^2).~~~~~ \label{b1y}
\end{eqnarray}
Figures (\ref{fig1}a) and (\ref{fig1}b) provide respectively the evolution of the components of the magnetic field and of the spin. The choice of smooth polynomial ansatz for the reverse engineering protocol generates a smooth solution. As intuitively expected, the shorter $t_f$, the larger the variation. This feature can be seen directly on Eqs.~(\ref{b1x}) and (\ref{b1y}) through the $1/t_f$ factors.


In conclusion of this section, we have shown that different formulations of the same problem yield different class of solutions. Within a given formulation, there is an infinitely large number of solutions for given boundary conditions. Those observations are useful to setup protocols for which we will add more constraints. In the following, we discuss the simultaneous spin flip of two spin having different gyromagnetic factors and the design of magnetic field trajectories that ensure an optimal spin flip fidelity robust against the value of the exact value of the gyromagnetic factor.
We will focus on the precession equations which presents the advantage of a direct possible visualization of the spin trajectory on the Bloch sphere.

\section{Simultaneous control of two different spins}

\begin{figure}[h!]
\begin{center}
\scalebox{0.55}[0.5]{\includegraphics{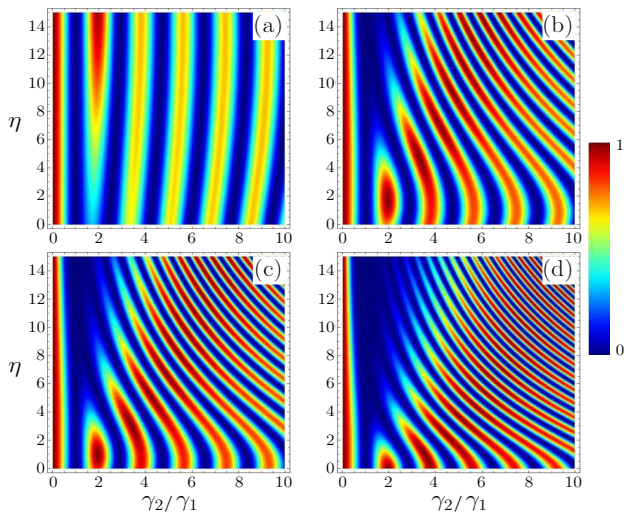}}
\caption{(color online) We design the magnetic field ${\bf B}(\eta,\kappa,t)$ components as a function of time to ensure the exact spin flip of spin 1 (of gyromagnetic factor $\gamma_1$) in an amount of time $t_f$. The parameter space of such solutions has two free extra parameters $\eta$ and $\kappa$. We then calculate the evolution of spin 2 in the time interval $[0,t_f)$ in the presence of ${\bf B}(\eta,\kappa,t)$. We plot the probability, $\Delta$, that spin 2 remains in its initial state as a function of $\gamma_2/\gamma_1$ and $\eta$ parameter for different values of the $\kappa$ parameter: (a) $\kappa=0.5$,  (b) $\kappa=2.5$, (c) $\kappa=3.1$ and (d) $\kappa=4.5$.}
\label{fig3}
\end{center}
\end{figure}

\begin{figure}[h!]
\begin{center}
\scalebox{0.5}[0.5]{\includegraphics{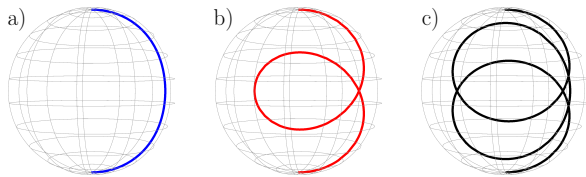}}
\caption{Example for which the same ${\bf B}(t)$ spin flips perfectly 3 different spin having different gyromagnetic factor: (a) $\gamma_1=2$, (b) $\gamma_2=5.34$ and (c) $\gamma_3=8.94$. Parameters: $\kappa=0.5$ and $\eta=5$. 
}
\label{fig4}
\end{center}
\end{figure}

\subsection{Spin flip of different spins}
We now consider a second spin ${\bf S}_2$ having a different gyromagnetic factor $\gamma_2$ (we assume that there is no interactions between the two spins). To setup the reverse engineering protocol allowing to control both spins with the \emph{same} time-dependent magnetic field, we proceed in the following manner: we enlarge the space of functions that flip the first spin, and search for the subset of parameters that also ensure the spin flip of the second spin. We shall use the same variation as previously for $\theta$ ($=\pi s$) but a more involved $\varphi(s)$ ansatz with two free parameters, $\kappa$ and $\eta$: $\varphi(s)=\kappa \left[ s + (\eta-1)s^2-2\eta s^3 + \eta s^4\right]$. This interpolating function fulfills the required boundary conditions $\varphi(0)=0$, $\varphi(1)=0$ and $\dot \varphi(1/2)=0$. Using Eqs.~(\ref{b1xn}) and (\ref{b1yn}), we can readily infer the time-dependent components of the magnetic field that one should apply.

 In Fig.~\ref{fig3}, we plot $\Delta=1-|\langle - | \psi(t_f)\rangle|^2=(1+S_{2z})/2$ as a function of the two parameters $\gamma_2/\gamma_1$ and $\eta$, and this for different values of $\kappa$. $\Delta$ provides a direct measurement of the projection of the spin on the $z$ axis at the end of the transformation. The blue zone are those for which we approach the target of a perfect reversing of spin two. This calculation shows (i) that whatever is the ratio $\gamma_2/\gamma_1$ there exists a couple of $(\eta,\kappa)$ parameters that will ensure a perfect rotation of the two spins despite the fact that their coupling strength to the magnetic field is different and (ii) the existence of dense blue zones (for $\gamma_2 \sim \gamma_1$) for which the rotation for both spin can be very good, this feature is the one required for robustness against dispersion of the values of $\gamma_2$ (see below). Actually, the existence of many curves with minimum values of $\Delta$ in Fig.~\ref{fig3} means that we can simultaneously spin flip many spins having different gyromagnetic factors with the appropriate magnetic field. An example is depicted in Fig.~\ref{fig4} for three different spins where we have represented on the Bloch sphere the time-evolution of each spin. Interestingly, our protocol generates loops on the Bloch sphere to ensure that all spin trajectories end up at the opposite pole at the same time. The one loop trajectory is reminiscent of the spin echo technique but is here generated automatically by our protocol.

 \subsection{Magnetic field shaping to ensure the robustness of the spin flip protocol}

The reverse engineering protocol is well adapted to add further constraints. An important issue is to design spin flips protocols that are robust against the dispersion in the parameters governing the time evolution of the system. A standard example is provided by the dispersion of Larmor frequencies of an ensemble of two-level systems in liquid and solid NMR experiments \cite{nmr}. This question is important for the implementation of quantum computing algorithm \cite{lukin,mohan}.

\begin{figure}[h!]
\begin{center}
\scalebox{0.4}[0.4]{\includegraphics{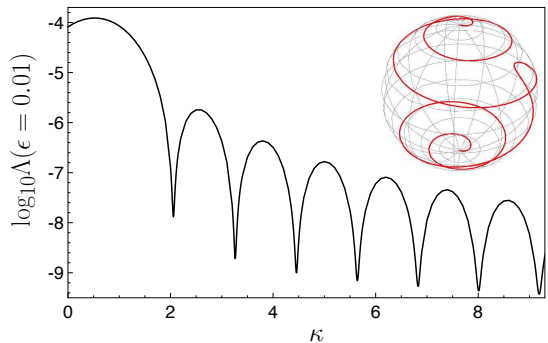}}
\caption{Robustness function $\log_{10}[\Lambda(\epsilon=0.01)]$, as a function of $\kappa$ for $\eta=20$. Inset: Spin flip trajectory of the second spin on the Bloch sphere with $\gamma_2=1.01\gamma_1$, $\eta=20$ and $\kappa=9.18918$.  For this large magic value of $\kappa$, we get $\log_{10} [ \Lambda(\epsilon=0.01) ]=-9.4297$.}
\label{fig5}
\end{center}
\end{figure}

To address this issue, we introduce  $\Lambda(\epsilon)$ which measures an average distance towards the exact spin flip by averaging the different probabilities of remaining in the initial state in an interval of size $2\bar \gamma\epsilon$ about the mean gyromagnetic factor $\bar \gamma$ under consideration:
\begin{equation}
\Lambda(\epsilon)=\frac{1}{2 \bar \gamma \epsilon}\int_{\bar \gamma(1-\epsilon)}^{\bar \gamma(1+\epsilon)} \Delta(\gamma_2)d\gamma_2.
\label{lambda}
\end{equation}
Figure \ref{fig5} shows the decimal logarithm of the robustness function $\Lambda$  as a function of $\kappa$ where $\epsilon=0.01$ and $\eta=20$ is fixed. We observe the existence of a set of discrete ``magic'' values for $\kappa$ that ensures an optimal spin flips. The quality of the spin flips increases with the value of the magic $\kappa$ value. For instance, we get $\log_{10}[\Lambda(\epsilon=0.01)] = -7.8816$ for the first magic value $\kappa = 2.0564$, $\log_{10}[\Lambda(\epsilon=0.01)] = -8.7127$ for $\kappa = 3.262$. We have also represented the evolution of the second spin on the Bloch sphere in the inset of Fig.~\ref{fig5} for $\kappa=9.1892$, which corresponds to $\log_{10} [ \Lambda(\epsilon=0.01) ]=-9.4297$.

For a given time duration $t_f$ of the process, the robustness increases at the expense of an increasingly large transient magnetic field amplitude. The use of high optimal values of magic $\kappa$ generates many rotations of the spin on the Bloch sphere (see the inset of Fig.~\ref{fig5}). This is not surprising since it simply generalizes somehow the spin echo technique.

\begin{figure}[h!]
\begin{center}
\scalebox{0.4}[0.4]{\includegraphics{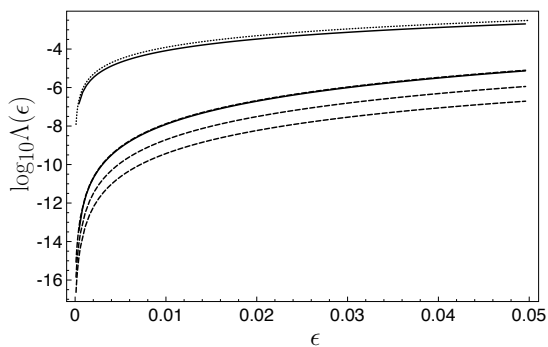}}
\caption{Robustness function of $\log_{10}[\Lambda(\epsilon)]$ as a function of $\epsilon$ for different spin flip protocols: $\pi$-pulse (upper solid line), spin echo technique (lower solid line), reverse engineering protocol for the non magic value $\kappa=0.5$ (dotted line), reverse engineering protocols for the magic values $\kappa= 2.0564$, $\kappa= 3.262$, and $\kappa=9.1892$ (and dashed lines). The larger the magic value the lower the robustness function.}
\label{fig6}
\end{center}
\end{figure}

Figure \ref{fig6} summarizes the robustness functions of different spin flip protocols for gyromagnetic factors spanning the interval $2\bar \gamma\epsilon$ about the mean value $\bar \gamma$. It compares the performance of (i) the simple $\pi-$pulse designed for the mean value $\bar \gamma$ and whose explicit expression is derived in Appendix, (ii) the spin echo technique (see Appendix) and different reverse engineering protocols. We include those latter protocols for a non magic value of the $\kappa$ parameter and for three magic values. The first magic value is already competitive with the spin echo technique (nearly superposition of the two corresponding robustness function). The larger magic values clearly improve efficiently the fidelity of the spin flip operation on the whole interval investigated here (up to 5\% difference of the gyromagnetic factor).

\subsection{Simultaneous perfect spin flip and superposition state generation}

\begin{figure}[h!]
\begin{center}
\scalebox{0.6}[0.6]{\includegraphics{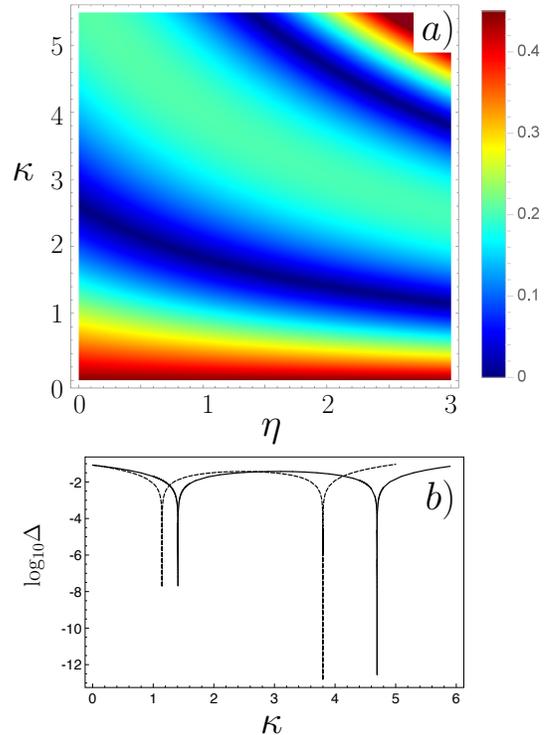}}
\caption{a) Mean value of the $z$ component of the second spin at $t_f$ (in decimal logarithmic scale) as a function of $\eta$ and $\kappa$ with $\gamma_1=2$ and $\gamma_2=1$. b) Cut section of Fig.~\ref{fig8} as a function of $\kappa$ for $\eta=2$ (solid line) and $\eta=3$ (dashed line).}
\label{fig8}
\end{center}
\end{figure}

The method presented in the previous sections can be readily generalized for other requirements. One could spin flip spin 1 and require for the spin 2 to end up in the horizontal plane of the Bloch sphere (superposition of state up and down with the same weight).

To confirm such possibilities offered by our extended space of parameter, we fix the ratio $\gamma_2/\gamma_1$, and plot the new $\Delta$ ($=S_{2z}$) function as a function of both parameters $\eta$ and $\kappa$. An example is provided in Fig.~\ref{fig8}a for $\gamma_2/\gamma_1=0.5$. The two cut of the 2D plot (Fig.~\ref{fig8}b) taken for two values of the $\eta$ parameter shows explicitly the existence of two values of $\kappa$ that ensures an optimal spin flip of spin 1 and that rotate spin 2 in the equatorial plane of the Bloch sphere. For instance, the optimal parameters are here: for $\eta=2$, $\kappa=4.6936$ with $\log_{10}(S_{2z})=-12.5451$  and for $\eta=3$, $\kappa=3.799$ with $\log_{10}(S_{2z})=-12.8491$.

Our method is generic. For instance with $\gamma_1=3$ and $\gamma_2=1$, our protocol also provides an optimal solution for the same target states.
As an example, we find for $\eta = 8$, $\kappa=5.429$ with $\log_{10}(S_{2z})= -12.6294$.

\section{Control of the spin trajectory in the presence of interactions}

In this section, we extend the results presented in the previous sections  to the situation for which there is an isotropic mutual interactions between the two spins. Interestingly, the solution can be readily obtained from that without interactions. We then discuss a more involved strategy for the case of exchange interaction $V^{(dd)}_{\rm int}=4\xi s_{1z} .  s_{2z}$ also referred to as the Ising interaction \cite{vitanov01,bose04,Nam15}.

\subsection{Isotropic interactions}

Consider a magnetic field function ${\bf B}_0(t)$ that solves simultaneously the equations (\ref{eqs1}) for spin 1 of gyromagnetic factor $\gamma_1$ and spin 2 of gyromagnetic factor $\gamma_2$ for the desired target states. It obeys
\begin{eqnarray}
\partial_t {\bf S}^0_1  & = &  \gamma_1 {\bf B}_0(t) \times {\bf S}^0_1 , \nonumber \\
\partial_t {\bf S}^0_2  & = &  \gamma_2 {\bf B}_0(t) \times {\bf S}^0_2 .
\label{eqC1}
\end{eqnarray}

In the presence of isotropic interactions ($V^{(is)}_{\rm int}=\mu \bf s_1 \cdot \bf s_2$), we search the magnetic field function ${\bf B}(t)$ that one should apply to reach the same target states. We therefore have to solve
\begin{eqnarray}
\partial_t {\bf S}_1  & = &  \gamma_1 {\bf B}(t) \times {\bf S}_1 + \mu {\bf S}_2 \times {\bf S}_1, \nonumber \\
\partial_t {\bf S}_2  & = &  \gamma_2 {\bf B}(t) \times {\bf S}_2 + \mu {\bf S}_1 \times {\bf S}_2.
\label{eqC2}
\end{eqnarray}
Let's search for a solution of the form ${\bf B}(t)={\bf B}_0(t) + \alpha  {\bf S}_1 + \beta  {\bf S}_2$, where $\alpha$ and $\beta$ are two constant parameters that need to be determined. We have
\begin{eqnarray}
\partial_t {\bf S}_1  & = &  \gamma_1 {\bf B}_0(t) \times {\bf S}_1 + (\mu -\beta \gamma_1){\bf S}_2 \times {\bf S}_1, \nonumber \\
\partial_t {\bf S}_2  & = &  \gamma_2 {\bf B}_0(t) \times {\bf S}_2 + (\mu - \alpha \gamma_2){\bf S}_1 \times {\bf S}_2.
\label{eqC3}
\end{eqnarray}
Choosing $\beta=\mu/\gamma_1$ and $\alpha=\mu/\gamma_2$, we now have to solve the set of equations
\begin{eqnarray}
\partial_t {\bf S}_1  & = &  \gamma_1 {\bf B}_0(t) \times {\bf S}_1 , \nonumber \\
\partial_t {\bf S}_2  & = &  \gamma_2 {\bf B}_0(t) \times {\bf S}_2 .
\label{eqC4}
\end{eqnarray}
and we know the solution ${\bf S}_1={\bf S}_1^0$ and ${\bf S}_2={\bf S}_2^0$. We conclude that the system (\ref{eqC2}) admits the solution
${\bf S}_1={\bf S}_1^0$ and ${\bf S}_2={\bf S}_2^0$ with a magnetic field that varies as
\begin{equation}
{\bf B}(t)={\bf B}_0(t) + \frac{\mu}{\gamma_2}  {\bf S}_1^0 + \frac{\mu}{\gamma_1}  {\bf S}_2^0.
\end{equation}
In other words, once we have a solution for two independent spins (including in the case of two different gyromagnetic factors) we have also the solution for two spins that interact through an isotropic interaction potential of the form $V^{(is)}_{\rm int}=\mu \bf s_1 \cdot \bf s_2 $ whatever is the strength of interaction between the two spins. However, one can never rich the Bell state $|{\rm Bell}\rangle=(|+ -\rangle+|- +\rangle)/\sqrt{2}$ with isotropic interactions.

\subsection{Ising interactions}
\label{anis}

To reach such a state, one needs anisotropic interactions. The Hamiltonian of two identical spins 1/2 therefore reads
\begin{equation}
H = -\gamma {\bf S}_1\cdot {\bf B}(t)-\gamma {\bf S}_2\cdot {\bf B}(t)  +V^{(dd)}_{\rm int}.
\end{equation}
It is block diagonal in the basis that classifies the states by their angular momentum {$|++\rangle$, $|{\rm Bell}\rangle$ and $|A\rangle=(|+-\rangle - |-+\rangle)/\sqrt{2}$}.

As we are interested in the simultaneous spin flip of the two interacting spins or in the generation of the Bell state, we search for a solution in the subspace of angular momentum $J=1$:  $|\psi(t)\rangle = a(t)|++\rangle + b(t)|{\rm Bell}\rangle + c(t)|--\rangle$. The time dependent complex coefficients $a(t)$, $b(t)$ and $c(t)$ obey the set of linearly coupled equations
\begin{eqnarray}
i \hbar \dot a &=& a (\gamma B_z+\xi)+b B_- \gamma/\sqrt 2,   \label{eqa}\\
i \hbar \dot b &=& a B_+ \gamma/\sqrt 2-b \xi+c B_- \gamma/\sqrt 2, \label{eqb}\\
i \hbar \dot c &=& b B_+ \gamma/\sqrt 2+c (-\gamma B_z+\xi), \label{eqc}
\end{eqnarray}
where $\gamma$ is the gyromagnetic factor and $B_\pm = B_x \pm i B_y$.


The adiabatic passage techniques shows that the dynamics is amenable to a 2$\times 2$ submatrix involving only $a$ and $b$ variables \cite{PRL87}. The adiabaticity requires a transformation on a typical time scale of $30\hbar/\xi$. The shortcuts to adiabaticity techniques can be used in this subspace to accelerate the transition from the fully polarized state $|++\rangle$ to the Bell state $|{\rm Bell}\rangle$ \cite{bellstate}.

For this purpose, we search for a solution that corresponds to a transverse rotating magnetic field whose amplitude varies as a function of time: $\gamma B_x(t)=B(t) \cos(\omega t)$ and $\gamma B_y(t)=B(t) \sin(\omega t)$.
The sub matrix on $a$ and $b$ variables can be recast in a symmetric form within the interaction picture
\begin{equation}
H_I(t)=\left(
  \begin{array}{cc}
    \Delta(t)/2 & B(t)/\sqrt2 \\
    B(t)/\sqrt2 & -\Delta(t)/2 \\
  \end{array}
\right),
\end{equation}
where the diagonal time dependent coefficient is related to the longitudinal magnetic field component: $\Delta(t)=\gamma B_z(t)-\omega+2\xi$.
A convenient and classical method to implement reverse engineering in this context relies on the use of dynamical invariants. This method simply consists in determining a matrix $I(t)$ that fulfills the following relation
\begin{equation}
\frac{d I(t)}{dt}=\frac{\partial I(t)}{\partial t}-\frac{i}{\hbar} [I(t), H_I(t)]=0,
\label{eqinv}
\end{equation}
with the boundary conditions $[H_I(0),I(0)]=[H_I(t_f),I(t_f)]=0$. To find this matrix $I(t)$, we use the Lie algebra operators and expand $I(t)$ on the Pauli matrices: $I(t)={\bf u}\cdot \boldsymbol{\sigma}$ where ${\bf u}$ is a unit vector of spherical angles $(\theta,\varphi)$. The eigenstate of $I(t)$, $|\phi_+(t)\rangle = \cos(\theta/2) e^{i \varphi} |++\rangle+\sin (\theta/2) |{\rm Bell}\rangle$, is also the eigenstate of $H_I(t)$ for initial and final time according to the commutation relations for boundary conditions.
The reverse engineer method amounts to fixing the evolution of the vector ${\bf u}$ in order to interpolated the evolution between the initial state $|\phi_+(0)\rangle \propto |++\rangle$ and  the desired final state $|\phi_+(t_f)\rangle \propto |{\rm Bell}\rangle$. To this end, the commutation relations shall be transposed as boundary conditions for the time dependent variables $\theta(t)$ and $\varphi(t)$: $\theta(0)=0$, $\theta(t_f)=-\pi$ , $\dot{\theta}(0)=0$, $\dot{\theta}(t_f)=0$, $\varphi(0)= -\pi/2$, $\varphi(t_f)=-\pi/2$, $\varphi(t_f/2) = -\pi/2$, $\dot{\varphi}(0)= -\pi/t_f$, and $\dot{\varphi}(t_f)=\pi/t_f$. From Eq.~(\ref{eqinv}), we obtain
the relation between the Hamiltonian variables and the angles of ${\bf u}$, namely,
\begin{eqnarray}
\dot{\theta} &=& \sqrt{2} B(t) \sin\varphi,\\
\dot{\varphi} &=& - \Delta(t) + \frac{\dot \theta}{\tan \theta \tan \varphi}.
\end{eqnarray}
We infer the value of $B(t)$ and $B_z(t)$, using the polynomial interpolation of minimum order according to the boundary conditions for the $\theta(t)$ and $\varphi(t)$ variables.

\begin{figure}
\centering
\includegraphics[width=7.5cm]{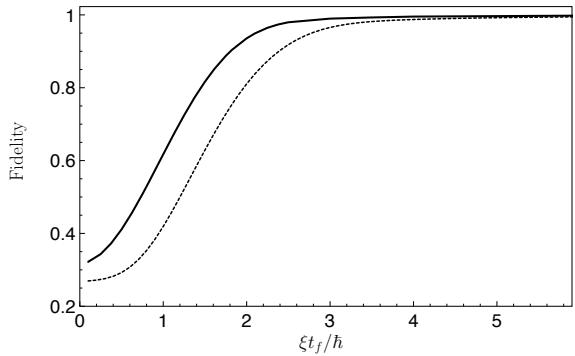}
\caption{Fidelity to reach the Bell state from a doubly polarized initial state $|++\rangle$ (i.e. $|\langle {\rm Bell}|\psi(t_f)\rangle |^2$) as a function of the time duration of the transformation  in the case of dipole-dipole interaction (solid line). Similar plot (i.e. $|\langle {\rm W}|\psi(t_f)\rangle |^2$) for three spins at the vertex of an equilateral triangle (dashed line).}
\label{figfidelity}
\end{figure}

In Fig.~\ref{figfidelity}, we plot the fidelity towards the desired Bell state at final time by solving the set of Eqs. (\ref{eqa})-(\ref{eqc}) with the magnetic field derived from the preceding approach. We find an improvement of at least one order of magnitude on the time required to reach the Bell state with high fidelity compared to the adiabatic evolution. As intuitively expected, it is impossible to accelerate to arbitrary short time. This is due to the finite value of the coupling (the $\xi$ parameter) that we kept fixed and which imposes a timescale $\hbar/\xi$.

Interestingly, this method can be readily generalized to more spins in a symmetric configuration. For instance, with three spins at the vertex of an equilateral triangle the same approach enables one to design in time the required fast evolution of the magnetic field components to drive the system from the fully polarized state $|+++\rangle$ to the ${\rm W}$ entangled state $=(|-++\rangle+|+-+\rangle+|++-\rangle)/\sqrt{3}$ (see dashed line in Fig.~\ref{figfidelity}).

\section{Conclusion}
In summary, we have proposed a reverse engineering approach to shape a time-dependent magnetic field to manipulate a single spin, two spins with different gyromagnetic factors, and two or more interacting spins in short amount of times. These techniques, as extension of previous STA techniques for atomic transport \cite{PRA2011,PRA2014,PRA2016NL}, provide robust protocols against the exact knowledge of the gyromagnetic factors for the one spin problem, or can be used to generate entangled states of two or more coupled spins. The analytical and smooth magnetic fields derived from reverse engineering are experimentally implementable, and the further optimization does not
requires time-dependent perturbation theory or numerical iteration, as compared to the previous results in Refs. \cite{2012njp,Tannor}.

Finally, we emphasize that the reverse engineering for spin dynamics provides powerful and effective language to implement the possible coherent control for
spin qubits by shaping time-dependent magnetic field. Since the spin 1/2 systems, and equivalent two-level systems, are ubiquitous in the  areas of quantum optics, the results, including fast and robust spin flip and entanglement generation, are applicable to quantum commutating and quantum information transfer, encompassing
rather different quantum systems, such as for example, two-coupled semiconductor quantum dots and cold atoms (or BECs) in double wells..

\section*{Acknowledgments}
This work was partially supported by the NSFC (11474193),
the Shuguang Program (14SG35), the Specialized Research Fund for the Doctoral Program (2013310811003),
the Program for Eastern Scholar,  the grant NEXT ANR-10-LABX-0037 in the framework of the Programme des Investissements
dAvenir and the Institut Universitaire de France. Q. Z. acknowledges the CSC PhD exchange program (grant number 201606890055).

\begin{appendix}
\section{Robustness function for the $\pi$-pulse and spin echo pulse sequence}

Consider a pulse of constant and homogeneous magnetic field $B_x$ along the $x$ direction of duration $\tau$.
The Hamiltonian of the spin $1/2$ particle that experience this field reads: $H_x=-{\bf M}\cdot {\bf B}$ with ${\bf M}=\gamma {\bf S}$ where $\gamma$ is the gyromagnetic factor. We have therefore $H=\Omega_x \hbar \sigma_x/2$  with $\Omega_x=-\gamma B_x$.  The propagator $U_x(\tau,0)$ which relates the initial and final states, $|\psi(\tau) \rangle = U_x(\tau,0)|\psi(0) \rangle$, reads
\begin{equation}
U(\tau,0) = \exp \left( \frac{\Omega_x  \sigma_x \tau}{2} \right) =  \cos \left( \frac{\omega_x \tau}{2} \right) -i  \sin \left( \frac{\omega_x \tau}{2} \right)\sigma_x.
\nonumber
\end{equation}
If the spin is initially in the state $|\psi(0) \rangle = | + \rangle$, a $\pi$-pulse spin flips the spin $|\psi(\tau) \rangle = | -\rangle$ corresponds to $\Omega_x \tau=\pi$. We note the corresponding propagator $U^{\pi}_x$. To estimate the robustness of the $\pi$-pulse, we consider such a pulse for a spin of gyromagnetic factor $\gamma_0$ and apply it to a spin of gyromagnetic factor $\gamma=\gamma_0(1-\epsilon)$. The probability that this latter spin has not spin flip is given by
$P_{++}(\epsilon)=|\langle + |U^{\pi}_x|+\rangle|^2= \sin^2\left( \frac{\pi \epsilon}{2} \right).$
The corresponding robustness function
$$
\Lambda^\pi(\epsilon)=\frac{1}{2\epsilon}\int_{-\epsilon}^\epsilon P_{++}(\epsilon)d\epsilon=\frac{1}{2}-\frac{\sin(\pi\epsilon)}{2\pi \epsilon} \simeq \frac{\pi^2\epsilon^2}{12}.
$$

The spin echo protocol corresponds to the propagator $U^{se}=U^{\pi/2}_xU^{\pi}_yU^{\pi/2}_x$.  The probability that the spin remains in its intial state $|+\rangle$ is $\tilde P_{++}(\epsilon)=|\langle + |U^{se}|+\rangle|^2=P^2_{++}(\epsilon)$, which results in the robustness function,
$$
\Lambda^{se}(\epsilon)= \frac{1}{2\epsilon}\int_{-\epsilon}^\epsilon \tilde P_{++}(\epsilon)d\epsilon=\frac{3}{8}-\frac{\sin(\pi\epsilon)}{2\pi\epsilon}+\frac{\sin(2\pi\epsilon)}{16\pi\epsilon} \simeq \frac{\pi^4\epsilon^4}{80}.
$$
\end{appendix}


\end{document}